\documentclass[twocolumn,secnumarabic,amssymb, amsmath, nofootinbib,tightenlines,
nobibnotes, aps, prl,epsfig]{revtex4}
\usepackage{graphicx}
\usepackage{dcolumn}
\usepackage{bm}
\begin{document}
\preprint{APS/123-QED}
\title{The ratio of the beauty structure functions $R^{b}=\frac{F_{L}^{b}}{F_{2}^{b}}$ at low-$x$ }

\author{G.R.Boroun}%
 \email{grboroun@gmail.com; boroun@razi.ac.ir }
\affiliation{ Physics Department, Razi University, Kermanshah
67149, Iran}
\date{\today}

\begin{abstract}
{ We study  the structure functions $F_{k}^{b}(x,Q^{2})$ ($k=2,
L$) and the reduced cross section $\sigma_{r}^{b}(x,Q^{2})$ for
small values of Bjorken$^{,}$s $x$ variable with respect to the
hard (Lipatov) pomeron for the gluon distribution and provide a
compact formula for the ratio $R^{b}$ that is useful to extract
the
 beauty structure function from the beauty reduced  cross section, in particular at DESY HERA.
Also we show that the effects of the nonlinear corrections to the
gluon distribution  tame the behavior of the beauty structure
function and the beauty reduced cross section at low $x$.}
\end{abstract}
\maketitle
\section{1. INTRODUCTION}

The measurement of the inclusive beauty  quark (b-quark) cross
section and the derived structure function $F_{2}^{b\overline{b}}$
in DIS at HERA is an important test of the theory of the strong
interaction, quantum chromodynamics (QCD), within the Standard
Model. Precise knowledge of the parton density functions (PDFs) is
for example essential at the Large Hadron Collider (LHC). The b-
quark density is important in Higgs production at the LHC in both
the Standard Model and in extensions to the Standard Model such as
supersymmetric models at high values of the mixing parameter
$\tan\beta$ [1]. First measurements [2] of the b-quark cross
sections at HERA were significantly higher than the QCD
predictions calculated at next-to-leading order (NLO)
approximation. The theoretical NLO QCD predictions are more than
three standard deviations below the experimental data. At the
Tevatron, old analyses indicated that the overall description of
the data can be improved [3] by adopting the non-perturbative
fragmentation function of the b-quark into the B-meson: an
appropriate treatment of the b-quark fragmentation properties
considerably reduces the disagreement between measured beauty
cross section and the results of corresponding NLO QCD
calculations. Also the latest measurement [4] of beauty
photoproduction at HERA is in a reasonable agreement with the NLO
QCD predictions or somewhat higher. This measurement  of the
beauty contribution to the inclusive proton structure function
$F_{2}(x,Q^{2})$ have been presented for small values of the
Bjorken scaling variable $x$, namely $1E{-4} < x < 6E{-2}$, and
for moderate and high values of the photon virtuality $Q^{2}$,
namely $5 \leq Q^{2} \leq 2000 ~GeV^{2}$. These data were based on
a dataset with an integrated luminosity of $189 ~pb^{-1}$, which
was about three times greater than in the previous measurements.
The data was recorded in the years 2006 and 2007 with $54~
pb^{-1}$ taken in e-p mode and $135 ~pb^{-1}$ in $e^{+}p$ mode.
The e-p center of mass energy is $\sqrt{s} = 319~ GeV$, with a
proton beam energy of $E_{p} = 920~ GeV$ and electron beam energy
of $E_{e} = 27.6~ GeV$
[4].\\
In the one-photon exchange approximation, beauty meson production
in deeply inelastic $ep$ scattering is via the photon-gluon fusion
subprocess $\gamma^{\star}+g{\rightarrow}b+\bar{b}$  and therefore
sensitive to the gluon density in a proton $f_{g}(x, \mu^{2})$.
The beauty structure function $F_{2}^{b\overline{b}}$  is obtained
from the measured  beauty-meson  cross section after applying
small correction for the longitudinal structure function
$F_{L}^{b\overline{b}}$ [5]. In the framework of DGLAP dynamics
[6], the leading contribution to the beauty meson production is
given by  two basic methods. One of them is based on the massless
evolution of parton distributions and the other one is based on
the massive boson-gluon fusion matrix element convoluted with the
gluon density of the proton [7-17]. The reduced cross section for
beauty quark introduced by H1 Collaboration
\begin{eqnarray}
\sigma^{b\overline{b}}_r&=&\frac{Q^4 x}{2\pi{\alpha}^2
Y_+}\frac{d^2\sigma^{b\overline{b}}}{dxdy}\nonumber\\
&&=F^{^{b\overline{b}}}_2
(x,Q^2,m^{2})-{\frac{y^2}{Y_+}}F^{^{b\overline{b}}}_L
(x,Q^2,m^{2})\nonumber\\
 &&=F^{^{b\overline{b}}}_2
(x,Q^2,m^{2})(1-{\frac{y^2}{Y_+}}R^{b}),
\end{eqnarray}
where in earlier HERA analysis, $F_{L}^{b\overline{b}}$ was taken
to be zero for simplicity. But for the NLO analysis the
$F_{L}^{b\overline{b}}$ contribution was subtracted from data.
Therefore determination of the longitudinal beauty structure
function at low $x$ at HERA is important because the
$F^{^{b\overline{b}}}_L$ contribution to the beauty cross section
can be sizeable. Here, we use the  hard (Lipatov)Pomeron behavior
of structure functions  and determine the ratio of the beauty
structure functions
$R^{b}=\frac{F^{^{b\overline{b}}}_L}{F^{^{b\overline{b}}}_2}$ from
this behavior in the limit of low $x$. The low-$x$ asymptotic
behavior of the parton densities are given by
\begin{eqnarray}
f(x,Q^{2})|_{x{\rightarrow}0}{\rightarrow}1/x^{1+\delta},
\end{eqnarray}
where $\delta$ is corresponding to the hard (Lipatov) Pomeron
intercept [18-20]. In the low-$x$ range, the gluon and
quark-singlet contributions are matter while the non-singlet
contributions are small. However, our analysis shows that the
predictions for $R^{b}$ with hard Pomeron intercept describe the
inclusive structure function with good accuracy to NLO at low-$x$,
and this analysis is independent
of the DGLAP evolution of the gluon distribution function.\\
In the present letter we use the hard-Pomeron  method to derive
the ratio of the beauty structure function and the reduced beauty
cross section in the region $Q^{2}>m^{2}$ at low-$x$. Section 2 is
devoted to the numerical solution of the master equation for the
beauty structure functions using gluon distribution
parameterizations. Then we consider the nonlinear corrections to
the gluon distribution function with respect to the GLR-MQ
equation for the $F_{2}^{b}(x,Q^{2})$
 and $\sigma_{r}^{b}(x,Q^{2})$. Our results and conclusions are given in
 sect.3.\\

\section{2. The method }

In the twist-2 approximation, for the beauty flavor case, the
beauty structure functions $F^{b\overline{b}}_{i}(x,Q^{2}),
(i=2,L)$ are described by
\begin{eqnarray}
F_{i}^{b}(x,Q^{2};m_{b}^{2})=\sum_{j}C_{i}^{j}(a_{s},x,\frac{Q^{2}}{\mu^{2}})
{\otimes}f_{j}(x,\mu^{2}),
\end{eqnarray}
where $a_{s}=\frac{\alpha_{s}(\mu^{2})}{4\pi}$, $f_{j}(x,\mu^{2})$
are the parton densities and
$C_{i}^{j}(a_{s},x,\frac{Q^{2}}{\mu^{2}})$ are the coefficients
functions. Here the mass factorization scale $\mu$ is assumed to
be either $\mu^{2}=4m_{b}^{2}$ or $\mu^{2}=4m_{b}^{2}+Q^{2}$, and
the symbol ${\otimes}$ for heavy flavour structure functions
denotes convolution according to the integral
\begin{eqnarray}
[A{\otimes}B](x)=\int_{\zeta}^{1}\frac{dz}{z}A(z)B(\frac{x}{z}),
\end{eqnarray}
 where $\zeta = x(1 + 4\xi)$ $(\xi{\equiv}\frac{m^{2}_{b}}{Q^{2}})$. In the $\overline{MS}$ scheme, the coefficient functions $C_{i}^{j}$
depend on the scaling variables $\eta(=(s -
4m^{2}_{b})/4m^{2}_{b})$ and $\xi$, where $s$ is the square of the
center of mass energy of the virtual photon-parton subprocess
$Q^{2}(1 - z)/z$ and $z$ is the fraction of energy carrying by the
parton as
\begin{equation}
z=\frac{Q^{2}}{2q.p_{p}},\nonumber
\end{equation}
where $p_{p}$ is the partonic four-momentum transfer in
boson-parton fusion processes. Based on the hard-Pomeron behavior
for the parton densities at low-$x$ (Eq.2), we  determine the
solutions of the beauty structure functions in Eq.3. After the
conventional integration (according to Eq.4) and doing some
rearranging, Eq.3 can be rewritten as
\begin{eqnarray}
F_{i}^{b}(x,Q^{2};m_{b}^{2})=\sum_{j=g,S}xf_{j}(x,\mu^{2})[C_{i}^{j}(a_{s},x,\frac{Q^{2}}{\mu^{2}})
{\otimes}z^{\delta}],
\end{eqnarray}
with hard behavior for the intercept at Regge factorization, where $[....]=\int_{x}^{\frac{1}{a}}C_{i}^{j}(..,z)z^{\delta}dz$ and $a=1 + 4\xi$.\\
Thus, considering structure functions with respect to hard
intercept are given by the following expression
\begin{eqnarray}
F_{i}^{b}(x,Q^{2};m_{b}^{2})&=&xf_{g}(x,\mu^{2})[C_{i}^{g}(a_{s},x,\frac{Q^{2}}{\mu^{2}})
{\otimes}z^{\delta}]\nonumber\\
&&+xf_{S}(x,\mu^{2})[C_{i}^{S}(a_{s},x,\frac{Q^{2}}{\mu^{2}})
{\otimes}z^{\delta}].
\end{eqnarray}
Here the coefficient functions are defined with respect to their
origin. The coefficient function $C^{g}$ originate from the
partonic subprocesses where the virtual photon is coupled to the
heavy quark, whereas the $C^{S}$ comes from the subprocess where
the virtual photon interacts with the light quark. The lowest
order term contains only the gluon density so that its
proportional to $C^{g}$. Whereas the light quark densities only
come in next order analysis and those is  about $5\%$. Then
$F_{i}^{b}$$^{,}$s are used in global analyses  constrain the
gluon density. Besides the gluon density, the main source of
theoretical uncertainty in $F_{i}^{b}$$^{,}$s is the values of the
beauty renormalization scales. Therefore, a further simplification
is obtained by neglecting the $\gamma^{*}q(\overline{q})$ fusion
subprocesses in Eq.6, which is justified because their
contributions vanish at LO and are small at NLO for small values
of $x$ [21]. Thus, we derive the low-$x$ approximation formula for
the beauty structure functions at LO up to NLO by the following
form
\begin{eqnarray}
F_{i}^{b}(x,Q^{2};m_{b}^{2}){\simeq}xf_{g}(x,\mu^{2})[C_{i}^{g}(a_{s},x,\frac{Q^{2}}{\mu^{2}})
{\otimes}z^{\delta}].
\end{eqnarray}
Our prediction for the ratio of the beauty structure functions
$R^{b}$ is independent of the gluon density at low-$x$, as
\begin{eqnarray}
R^{b}&=&\frac{F^{b}_L}{F^{b}_2}\nonumber\\
&&=\frac{[C_{L}^{g}(a_{s},x,\frac{Q^{2}}{\mu^{2}})
{\otimes}z^{\delta}]}{[C_{2}^{g}(a_{s},x,\frac{Q^{2}}{\mu^{2}})
{\otimes}z^{\delta}]}.
\end{eqnarray}
In the above expression $C^{g}_{j}$$^{,}$s are the gluonic
coefficient functions expressed in terms of LO and NLO
contributions in beauty-quark leptoproduction  as follows
\begin{eqnarray}
C^{g}_{j}{\rightarrow}C^{0}_{j,g}+a_{s}(\mu^{2})[C^{1}_{j,g}
+\overline{C}_{j,g}^{1}\ln\frac{\mu^{2}}{m_{b}^{2}}].
\end{eqnarray}
The coefficients $C^{0}_{j,g}$ and
$C^{1}_{j,g}(\overline{C}^{1}_{j,g})$ are at LO and NLO
respectively. These coefficient functions have been  computed at
LO  up to NLO in Refs. [7-9,22-27]. We conclude that the ratio of
the beauty structure functions (the Callan-Gross ratio) in
heavy-quark leptoproduction is a phenomenological observable
quantitatively defined in pQCD. With respect to the experimental
aspect, it is useful for extraction of the reduced beauty cross
section from the phenomenological relation for structure function
$F_{2}^{b}$ and the ratio $R^{b}$ as follows
\begin{eqnarray}
\sigma^{b\overline{b}}_r&=&xf_{g}(x,\mu^{2})[C_{2}^{g}(a_{s},x,\frac{Q^{2}}{\mu^{2}})
{\otimes}z^{\delta}]\nonumber\\
&&{\times}(1-{\frac{y^2}{Y_+}}\frac{[C_{L}^{g}(a_{s},x,\frac{Q^{2}}{\mu^{2}})
{\otimes}z^{\delta}]}{[C_{2}^{g}(a_{s},x,\frac{Q^{2}}{\mu^{2}})
{\otimes}z^{\delta}]}).
\end{eqnarray}
This  formula can reproduce the HERA results for the beauty
reduced cross section from the phenomenological models for the gluon distribution function at low-$x$.\\
At low-$x$, the high-density gluon distributions are singular for
$b\overline{b}$ meson production. The density of low
momentum-fraction gluons is expected to be close to saturation of
the available phase space, so as to produce significant
recombination effects. Therefore, the gluon distribution function
will be close to phase-space saturation and here will be important
gluon-fusion effects ($gg{\rightarrow} g$). These effects can be
accounted for the gluon scale evolution equation by adding a
negative nonlinear (quadratic) term to the standard linear DGLAP
term as
\begin{eqnarray}
\frac{{\partial}f_{g}(x,Q^{2})}{\partial \log
Q^{2}}&=&[\mathrm{DGLAP~ term ~of}~
\mathrm{order}(f_{g})]\nonumber\\
&&- [\mathrm{term~ of~ order}(f^{2}_{ g} )] .
\end{eqnarray}
Therefore, the linear evolution equation in this case is modified
by non-linear term description gluon recombination. An important
point in the gluon saturation approach is the $x$-dependent
saturation scale $Q_{s}^{2}(x)$. This scaling argument leads to
the conclusion that $\gamma^{*}p$ cross section, which is
$\textit{a priori}$ function of two independent variable ($x$ and
$Q^{2}$), is a function of only variable
$\tau=\frac{Q^{2}}{Q_{s}^{2}(x)}$. In the limit of high energy,
PQCD consistently predicts that the high gluon density should form
a Color Glass Condensate (CGC), where the interaction probability
in DIS becomes large and this is characterized by a hard
saturation scale $Q_{s}(x)$ which grows rapidity with $1/x$
[28-37]. In this region, the nonlinear saturation dynamics is
incorporated into the CGC model. As, it is valid only for $Q^{2}$
less than or of the order of the saturation momentum, which is at
most several $GeV^{2}$, while the fit result to SGK [31] model
extends up to $Q^{2}$ of the order of several hundred $GeV^{2}$.
 The overall physical picture is dependence to
the different regions in the ($x,Q^{2}$)-plane. For
$Q^{2}<Q_{s}^{2}(x)$ the linear evolution is strongly perturbed by
nonlinear effects where the parton system becomes dense and the
saturation corrections start to play an important role. In
color-dipole (CD) model the excitation of heavy flavors at low-$x$
is described in terms of interaction of small size quark-antiquark
$b\overline{b}$ color dipoles in the photon. This interaction can
be defined by the $\Psi_{L,T}^{b\overline{b}}(z,r)$, where it is
the probability to find  the $b\overline{b}$ color dipole in the
photon with respect to the $\sigma^{b\overline{b}}(x,Q^{2})$. Here
$z$ is the carrying fraction by the beauty quark into the photon's
light-cone momentum and $r$ is the size of the $b\overline{b}$
interaction in the photon. In this region the dipole cross section
is bounded by an energy independent value, as the dipole cross
section was proposed [28-37] to have the form
$\sigma_{dipole}(x,r)=\sigma_{0}\{1-exp(-r^{2}Q_{s}^{2}(x)/4)\}$
which impose the unitarity condition
($\sigma_{q\overline{q}}{\leq}\sigma_{0}$) for large dipole sizes
$r$. The CD cross section is given by color dipole factorization
formula
\begin{eqnarray}
\sigma^{b\overline{b}}(x,Q^{2})&=&\int_{0}^{1}\int
d^{2}\mathbf{r}[~|\Psi_{L}^{b\overline{b}}(z,\mathbf{r})|^{2}+|\Psi_{T}^{b\overline{b}}(z,\mathbf{r})|^{2}
]\nonumber\\
&&{\times}\sigma_{dipole}(x,\mathbf{r}).
\end{eqnarray}
 At small -$r$ region, the dipole cross section is related to
the gluon density where it is valid in the double logarithmic
approximation. As  the gluon density is
$xg(x,Q^{2}=Q_{s}^{2}(x))=r^{0}x^{-\lambda}$, and the parameter
$r_{0}$ specifies the normalization along the critical line. Thus,
the saturation scale is an intrinsic characteristic of a dense
gluon system. At low-$x$ region, the CD cross section satisfies
the solutions with Regge behavior for the beauty structure
function of the proton as
\begin{eqnarray}
F_{2}^{b}(x,Q^{2})=\frac{Q^{2}}{4\pi^2
\alpha_{em}}\sigma^{b\overline{b}}(x,Q^{2})=\sum_{n}f_{n}^{b}(Q^{2})(\frac{x_{0}}{x})^{\Delta_{n}}
\end{eqnarray}
where $n=soft,0,1,..$ and the choice $x_{0}=0.03$ [9-10,12]. The
intercepts $\Delta_{n}$ are equal to $\Delta_{0}/(1+n)$, where
$\Delta_{0}$ is defined with respect to the hard Lipatov pomeron
and the coefficients $f_{n}^{b}(Q^{2})$ are presented in Ref.12
(V.R.Zoller, Phys.Lett. B\textbf{509}, 69(2001)). The observation
of the QCD pomeron dynamics at distances $\sim m_{b}^{-1}$ leads
to exchange beauty exponent from $\Delta_{n}$ to $\Delta_{eff}$,
where $\Delta_{eff}$ is defined by
\begin{eqnarray}
\Delta_{eff}=\Delta_{0}[1-\sum_{n=1}r_{n}(1-\frac{\Delta_{n}}{\Delta_{0}})(\frac{x_{0}}{x})^{\Delta_{n}-\Delta_{0}}].
\end{eqnarray}

At what follows, we consider Eq.11 for the behavior of the beauty
structure functions as the recombination processes between gluons
in a dense system have to be taken into account and it has to be
tamed by screening effects. These nonlinear terms reduce the
growth of the gluon distribution in this kinematic region where
$\alpha_{s}$ is still small but the density of partons becomes so
large. Gribov, Levin, Ryskin, Mueller and Qiu (GLR-MQ)[38-39]
argued that the physical processes of interaction and
recombination of partons become important in the parton cascade at
a large value of the parton density, and that these shadowing
corrections could be expressed in a new evolution equation (the
GLRMQ equation).\\
Therefore, the evolution equations of gluons can be modified as
\begin{eqnarray}
\frac{{\partial}xf_{g}(x,Q^{2})}{{\partial}{\ln}Q^{2}}&=&\frac{{\partial}xf_{g}(x,Q^{2})}{{\partial}{\ln}Q^{2}}|_{DGLAP}\nonumber\\
&&-\frac{\alpha^{2}_{s}\gamma}{R^{2}Q^{2}}\int^{1}_{\chi}\frac{dy}{y}[yf_{g}(y,Q^{2})]^{2},
\end{eqnarray}
where the first term is the linear standard DGLAP evolution and
the second term is defined with respect to the 2-gluon density
[40-41]. In this equation, $\gamma$ is equal to $\frac{81}{16}$
for $N_{c}=3$ and  $R$ is the size of the target which the gluons
populate becomes so large that the annihilation of gluons becomes
important. $R$ will be of the order of the proton radius
$(R\simeq5\hspace{0.1cm} GeV^{-1})$ if the gluons are spread
throughout the entire nucleon, or much smaller
$(R\simeq2\hspace{0.1cm} GeV^{-1})$ if gluons are concentrated in
hot- spot [42] within the proton. Here  $\chi=\frac{x}{x_{0}}$
where $x_{0}(=0.01)$ is the boundary condition that the gluon
distribution joints smoothly onto the unshadowed region. In this
equation we need to determine the appropriate behavior of the
shadowing corrections to the gluon distribution at the initial
point. Where the unshadowed distribution to the gluon
($f_{g}^{u}$) at $Q_{0}^{2}$ has to modify for $x<x_{0}$ as
\begin{eqnarray}
xf_{g}(x,Q^{2}_{0})={xf_{g}}^{u}(x,Q^{2}_{0})[1+\theta(x_{0}-x)[{xf_{g}}^{u}(x,Q^{2}_{0})\nonumber\\
-{xf_{g}}^{u}(x_{0},Q^{2}_{0})]/{xf_{g}}^{sat}(x,Q^{2}_{0})]^{-1},
\end{eqnarray}
here
${xf_{g}}^{sat}(x,Q^{2})=\frac{16R^{2}Q^{2}}{27\pi\alpha_{s}(Q^{2})}$
is the value of the gluon which would saturate the unitarity limit
in the leading shadowing approximation. Equation (16) reduces to
the unshadowed form $xf_{g}^{u}$ when shadowing is negligible;
that is, when $xf_{g}^{sat}{\rightarrow}\infty$. In this region,
the interaction of gluons is negligible and we use the linear
evolution equations in $xf_{g}(x,Q^{2})$. However, at sufficiently
low-$x$, two gluons in different cascades may interact and so
decrease the gluon density. Therefore, we apply the saturation
corrections to the gluon distribution function in Eqs.7 and 10 to
obtain the shadowing corrections for the beauty structure function
and the beauty reduced cross section behavior and show that this
behavior tame at
low-$x$ limit.\\

\section{3. Results and Conclusions }

In this work, the beauty structure functions, the ratio of the
beauty structure functions and the reduced beauty cross sections
at $Q^{2}=$60 and 650 $GeV^{2}$ have determined. We compared our
results with H1 data on beauty production [4], color dipole model
[9-10,12] and the GJR parametrisation [43]. In Figs.1 and 2 the
beauty structure functions and reduced cross section with respect
to the gluon distribution function at the renormalization scale
$\mu^{2}=4m_{b}^{2}+Q^{2}$ with $m_{b}=4.57 ~GeV$ are shown. We
observe that these relations (Eqs. 7 and 10) are dependence to the
gluon distribution, which is usually taken from the GRV [22] and
Block [44] parameterizations or DL [18-20] model. In what follows
we shall use the gluon distribution with an intercept according to
the hard pomeron behavior at the DL model. These results show that
the fractional accuracy for the DL model with a hard pomeron
intercept is the best which confirms the correctness of our
solution with DL model when compared with H1 data , color dipole
model and  GJR parametrisation. For comparing with H1 data (2010)
[4], we presented our results from $Q^{2}=5~ GeV^{2}$ up to $2000~
GeV^{2}$ in Table 1. The theoretical uncertainties in our result
are according to the renormalization scales
$\mu^{2}=4m_{b}^{2}+Q^{2}$ and  $\mu^{2}=4m_{b}^{2}$. We observe
that the theoretical uncertainties, for the beauty functions,
related to the freedom in the choice of $\mu$ are negligibly
small. The agreement between our predictions with the results
obtained by H1 Collaboration  is remarkably good at the
renormalization scale
 $\mu^{2}=4m_{b}^{2}+Q^{2}$ and with the DL parton distribution
 function. As can be seen in all figures, the increase of our calculations for the beauty structure functions
towards low $x$ are consistent with the experimental data.\\
 In Fig.3  we
show the predicted ratio of the beauty structure functions,
$R^{b}$, as a function $Q^{2}$ at $x=0.001$ . This ratio is
independent of the choice of the gluon distribution function,
where approaches based on perturbative QCD and $\textit{k}_{T}$
factorization give similar predictions [13-15]. The $R^{b}$ effect
on the corresponding differential beauty cross section should be
considered in extraction of $F_{2}^{b}$. We see that this value is
approximately between 0.10 and 0.20 in a region of $Q^{2}$ and
this prediction for $R^{b}$ is close to the results Refs. [13-15].
We can see that the behavior of this ratio is in agreement with
the prediction from Ref.[8]. As authors obtained an approximate
formula at LO and NLO analysis for low-$x$ values, where at LO the
compact form is
\begin{eqnarray}
R^{b}{\approx}\frac{2}{1+4\xi}\frac{1+6\xi-4\xi(1+3\xi)J(\xi)}{1+2(1-\xi)J(\xi)},
\end{eqnarray}
where
$J(\xi)=-\frac{1}{\sqrt{a}}\ln(\frac{\sqrt{a}-1}{\sqrt{a}+1})$. We
compare our result with this compact formula in Fig.3.\\
In Figs. 4 and 5, the values of the nonlinear corrections to the
gluon distribution function determined for predictions of the
beauty structure functions  and beauty reduced cross section. In
Fig.4, we show shadowing corrections to the gluon distribution
function determined from Eq.(16) as a function of $x$ at the
initial scale $Q_{0}^{2}=5~GeV^{2}$ for determination of the
beauty structure function and beauty reduced cross section. In
Fig.5, we show the shadowing corrections to the gluon distribution
function determined from Eq.(15) at $Q^{2}=60~GeV^{2}$. We
observed that, as $x$ decreases, the singularity behavior of the
gluon functions are tamed by shadowing effects. Therefore the
singularity behavior of the beauty structure function and beauty
reduced cross section are tamed by shadowing effects. The solid
curves show the low-x behavior when shadowing is neglected, and
the lower curves (dash and dot) show the effect of the shadowing
contribution for R = 5 and 2 $GeV^{-1}$, respectively. We compared
our results with H1 data [4]
 and CD model [9-10,12] (dash-dot-dot). These results show that beauty structure function and beauty reduced cross section
 behaviors
are tamed with respect to nonlinear terms at the GLR-MQ equation
to the gluon density behavior at low $x$. These figures show that
screening effects are provided by multiple gluon interaction which
leads to the nonlinear terms in the DGLAP equation.\\

 In summary, our numerical results for the beauty structure functions and beauty reduced cross section at low x are obtained by
applying the hard (Lipatov) pomeron behavior at all $Q^{2}$ values
in the NLO analysis. We derived the
 ratio
 $R^{b}=\frac{F^{^{b\overline{b}}}_L}{F^{^{b\overline{b}}}_2}$ that is valid through NLO at small
  values of Bjorken$^{,}$s $x$ variable, as it is independent of the input gluon
 distribution function. To confirm the method and results, the calculated values are
compared with the H1 data and other models on the beauty structure
function, at low-$x$. Then we studied the effects of adding the
nonlinear GLR-MQ corrections to the  DGLAP evolution equation, by
adding to the beauty structure function and beauty cross section
at low-$x$. The nonlinear effects to the gluon distribution are
found to play an increasingly important at $x<0.001$. We observed
that, as $x$ decreases, the singularity behavior of the beauty
structure function and beauty reduced cross section are tamed by
shadowing effects. We compared our
results with the H1 data and with the CD model.\\
\acknowledgments G.R.Boroun thanks Prof.V.R.Zoller for discussions
and useful comments.
\newpage
\section{References}
1. D. Dicus, T. Stelzer, Z. Sullivan, and S. Willenbrock, Phys.
Rev. D{\textbf59}, 094016(1999);\\ C. S. Huang and S. H. Zhu,
Phys. Rev. D{\textbf60}, 075012(1999); \\C. Balazs, H. J. He, and
C. P. Yuan, Phys. Rev. D{\textbf60}, 114001(1999);\\ J. Campbell,
R. K. Ellis, F. Maltoni, and S. Willenbrock, Phys. Rev.
D{\textbf67}, 095002 (2003);\\ F. Maltoni, Z. Sullivan, and S.
Willenbrock, Phys. Rev. D{\textbf67}, 093005(2003).\\
2. C. Adloff et al. (H1 Collaboration), Phys. Lett. B
{\textbf467}, 156 (1999); Erratum: ibid B {\textbf518}, 331
(2001).\\
3. M. Cacciari and P. Nason, Phys. Rev. Lett.{\textbf 89},
122003(2002); M. Cacciari, S. Frixione, M.L. Mangano, P. Nason,
and G. Ridolfi, JHEP {\textbf0407}, 033(2004).\\
4. F.D. Aaron et al. [H1 Collaboration], Eur.Phys.J.C\textbf{65},
89 (2010); A.Aktas et al. [H1 Collaboration],
Eur.Phys.J.C\textbf{45}, 23 (2006); Eur.Phys.J.C\textbf{40}, 349
(2005).\\
5. A.V. Lipatov, N.P. Zotov, JHEP{\textbf0608}, 043(2006);
Phys.Rev.D {\textbf73}, 114018(2006).\\
6. Yu.L.Dokshitzer,
Sov.Phys.JETP {\textbf{46}}, 641(1977);\\
 G.Altarelli and
G.Parisi, Nucl.Phys.B \textbf{126}, 298(1977);\\
 V.N.Gribov and
L.N.Lipatov, Sov.J.Nucl.Phys. \textbf{15}, 438(1972).\\
7. J.Blimlein, A.De Freitas, W.L.van Neerven and S.Klein, Nucl.Phys.B\textbf{755}, 272(2006) .\\
8. A.~Y.~Illarionov,B.~A.~Kniehl and A.~V.~Kotikov, Phys.Lett. B {\bf 663}, 66 (2008).\\
9. N.Nikolaev, J.Speth and V.R.Zoller,Phys.Lett.B{\bf473}, 157(2000).\\
10. R.Fiore, N.Nikolaev and V.R.Zoller,JETP Lett{\bf90}, 319(2009).\\
11. I.P.Ivanov and N.Nikolaev,Phys.Rev.D{\bf65},054004(2002).\\
12. N.N.Nikolaev and V.R.Zoller, Phys.Lett. B\textbf{509},
283(2001); Phys.Atom.Nucl.\textbf{73}, 672(2010); V.R.Zoller,
Phys.Lett. B\textbf{509},
69(2001);\\
13. A.~V.~Kotikov, A.~V.~Lipatov, G.~Parente and N.~P.~Zotov Eur.\
Phys.\ J.\  C {\bf 26}, 51(2002).\\
14. V.~P.~Goncalves and M.~V.~T.~Machado, Phys.\ Rev.\ Lett.\ {\bf
91}, 202002 (2003).\\
15. N.Ya.Ivanov, Nucl.Phys.B\textbf{814}, 142(2009); Eur.Phys.J.C
{\bf59}, 647(2009).\\
16. G.R.Boroun and B.Rezaei,Nucl.Phys.B{\textbf857}, 143(2012);
EPL,{\textbf100}, 41001(2012); JETP, 2012, Vol.115,
No.3, 427(2012).\\
17. S.Catani, Z.Phys.C{\textbf75}, 665(1997).\\
18. A.Donnachie and P.V.Landshoff, Z.Phys.C \textbf{61},
139(1994); Phys.Lett.B \textbf{518}, 63(2001); Phys.Lett.B
\textbf{533}, 277(2002); Phys.Lett.B \textbf{470}, 243(1999);
Phys.Lett.B \textbf{550}, 160(2002).\\
19. R.D.Ball and P.V.landshoff, J.Phys.G\textbf{26}, 672(2000).\\
20. P.V.landshoff, arXiv:hep-ph/0203084 (2002).\\
21. E.Laenen, S.Riemersma, J.Smith and W.L. van Neerven,
Nucl.Phys.B \textbf{392}, 162(1993).\\
22. M.Gluk, E.Reya and A.Vogt, Z.Phys.C\textbf{67}, 433(1995);
Eur.Phys.J.C\textbf{5}, 461(1998). \\
23. V.N. Baier et al., Sov. Phys. JETP 23 104 (1966); V.G. Zima,
Yad. Fiz. 16 1051 (1972); V.M. Budnev et al., Phys. Rept. 15 181
(1974).\\
24. E. Witten, Nucl. Phys. B104 445 (1976); J.P. Leveille and T.J.
Weiler, Nucl. Phys. B147 147 (1979); V.A. Novikov et al., Nucl.
Phys. B136 125 (1978) 125.\\
25. A.Vogt, arXiv:hep-ph:9601352v2(1996).\\
26.  E.Laenen, S.Riemersma, J.Smith and W.L. van Neerven,
Nucl.Phys.B \textbf{392}, 162(1993).\\
27.  S. Catani, M. Ciafaloni and F. Hautmann, Preprint
CERN-Th.6398/92, in Proceeding of the Workshop on Physics at HERA
(Hamburg, 1991), Vol. 2., p. 690; S. Catani and F. Hautmann, Nucl.
Phys. B \textbf{427}, 475(1994); S. Riemersma, J. Smith and W. L.
van Neerven, Phys. Lett. B \textbf{347}, 143(1995).\\
28. K.Golec-Biernat, J.Phys.G {\bf28}, 1057(2002).\\
29. K.Golec-Biernat, Acta Phys.Pol.B {\bf35}, 3103(2004).\\
30. K.Golec-Biernat, Acta Phys.Pol.B {\bf33}, 2771(2002).\\
31. A.M.Stato, K.Golec-Biernat and J.Kwiecinski,
Phys.Rev.Lett.{\bf86}, 596(2001).\\
32. E.Iancu, K.Itakura and S.Munier, Phys.Lett.B {\bf590},
199(2004).\\
33.  E.Iancu, K.Itakura and S.Munier, Nucl.Phys.A {\bf708},
327(2002).\\
34. L. McLerran and R. Venugopalan, Phys.Rev.D\textbf{49},
2233(1994).\\
35. E. Iancu, A. Leonidov and L. McLerran,
Nucl.Phys.A\textbf{692}, 583(2001).\\
36.  E. Iancu, A. Leonidov and L. McLerran,
Phys.Lett.B\textbf{510},
133(2001).\\
37. E. Ferreiro, E. Iancu, A. Leonidov and L. McLerran, Nucl.Phys.A\textbf{703}, 489(2002).\\
38. A.H.Mueller and J.Qiu, Nucl.Phys.B\textbf{268}, 427(1986).\\
39. L.V.Gribov, E.M.Levin and M.G.Ryskin, Phys.Rep.\textbf{100},
 1(1983).\\
40. E.Laenen and E.Levin,Nucl.Phs.B.\textbf{451},207(1995).\\
41. K.J.Eskola, et.al., Nucl.Phys.B\textbf{660}, 211(2003).\\
42. E.M.Levin and M.G.Ryskin, Phys.Rep.\textbf{189}, 267(1990).\\
43. M. Gluck, P. Jimenez-Delgado, E. Reya, Eur.Phys.J.C
{\bf53},355 (2008).\\
44. M.M.Block et al., Phys.Rev.D
{\bf77}, 094003(2008).\\

\begin{figure}
\includegraphics[width=1\textwidth]{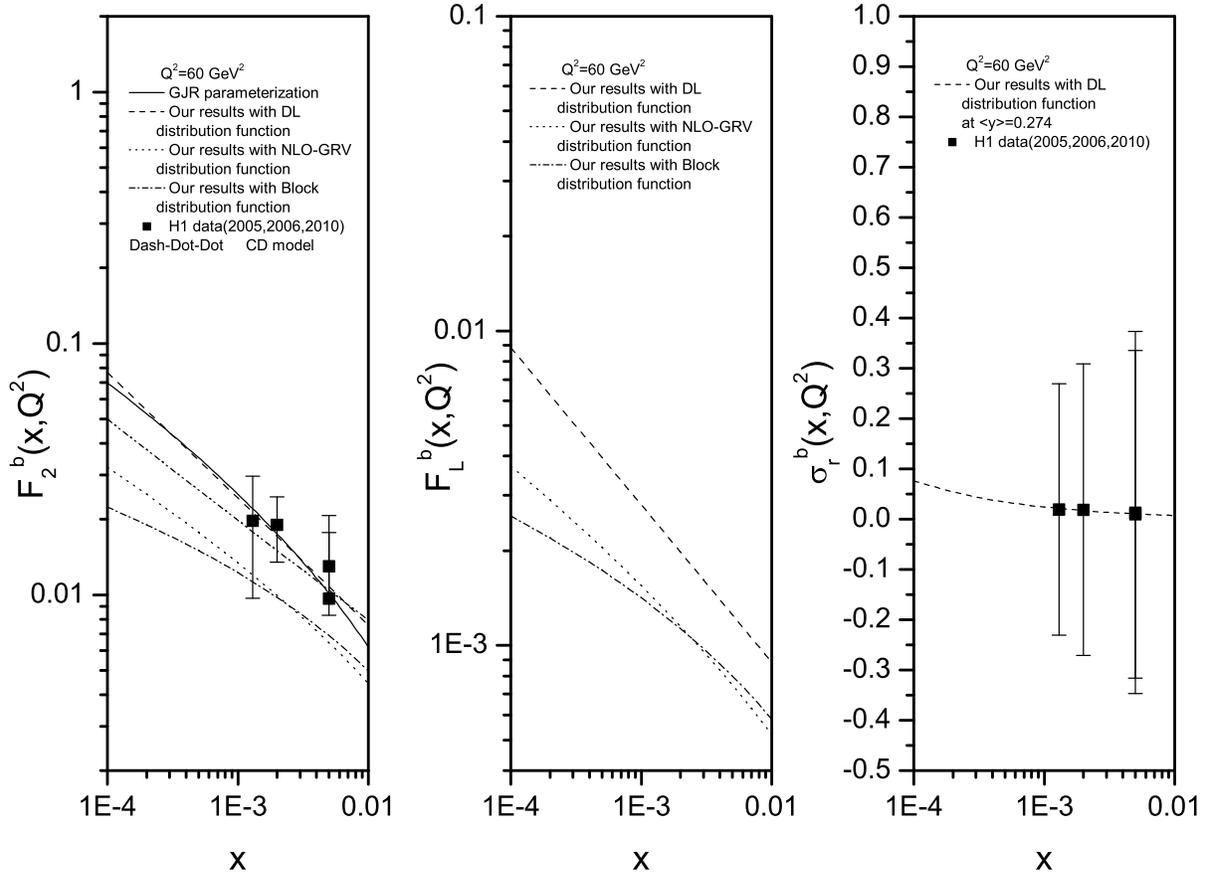}
\caption{Predictions for $F_{2}^{b}(x,Q^{2})$,
$F_{L}^{b}(x,Q^{2})$ and $\sigma_{r}^{b}(x,Q^{2})$ at $Q^{2}=60
GeV^{2}$ with the input gluon distribution from DL [18-20] model
(dash curve ), GRV [22] parameterization (dot curve) and Block
[44] model (dash-dot curve), compared with color dipole [9-10,12]
model (dash-dot-dot curve)
 and the GJR [43] parameterization (solid curve). }\label{Fig1}
\end{figure}
\begin{figure}
\includegraphics[width=1\textwidth]{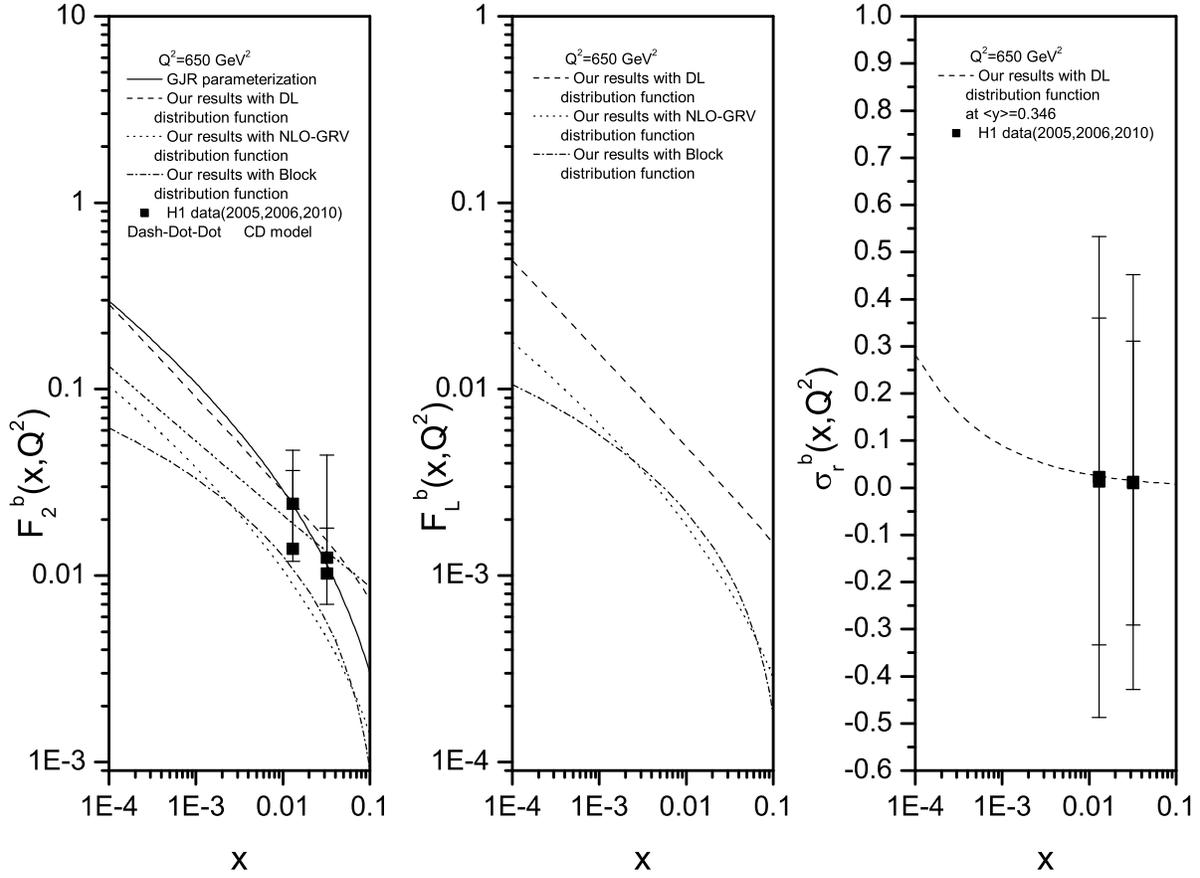}
\caption{The same as Fig.1 at $Q^{2} =650~GeV^{2}$. }\label{Fig1}
\end{figure}
\begin{figure}
\includegraphics[width=1\textwidth]{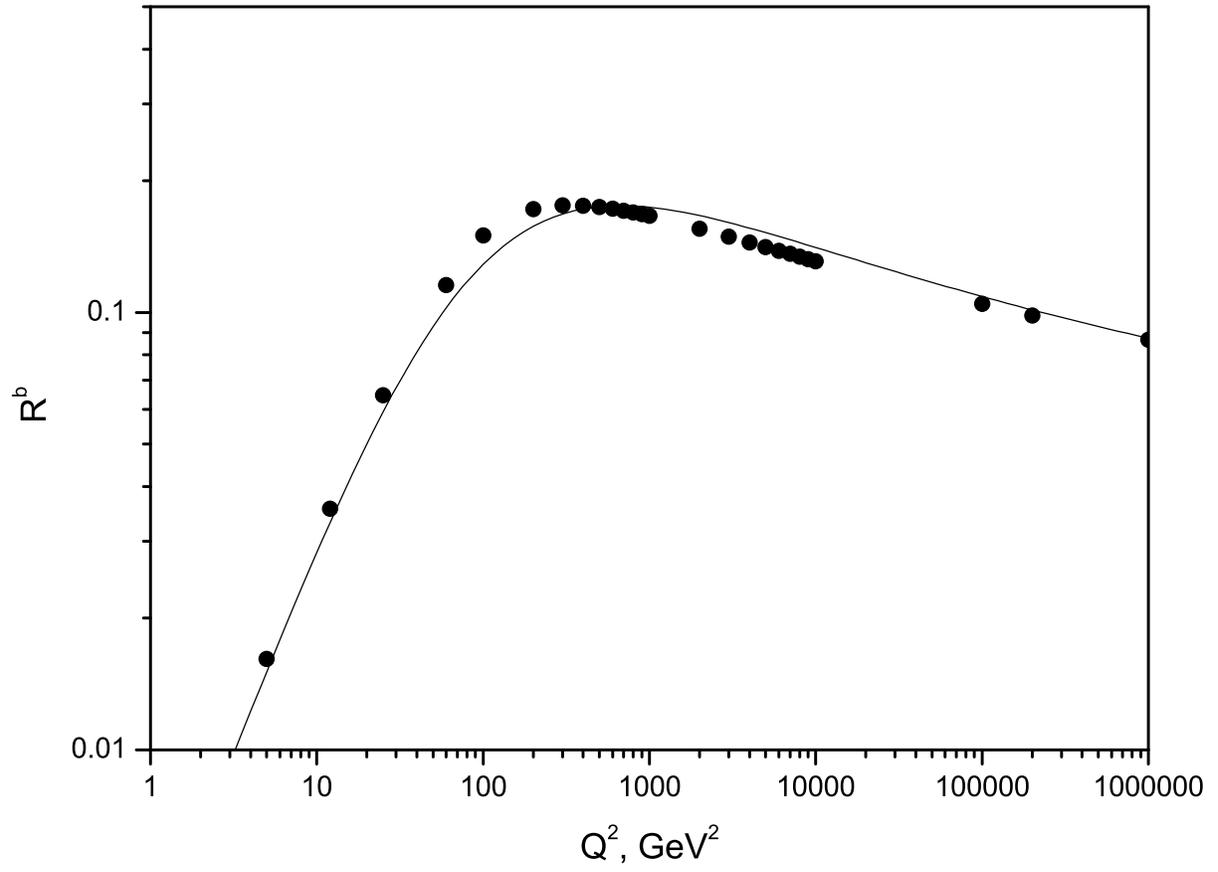}
\caption{$R^{b}$ evaluated as functions of $Q^{2}$ with
$\mu^{2}=Q^{2}+4m_{b}^{2}$ compared with LO compact formula in
Ref.[8](Solid curve). }\label{Fig1}
\end{figure}
\begin{figure}
\includegraphics[width=1\textwidth]{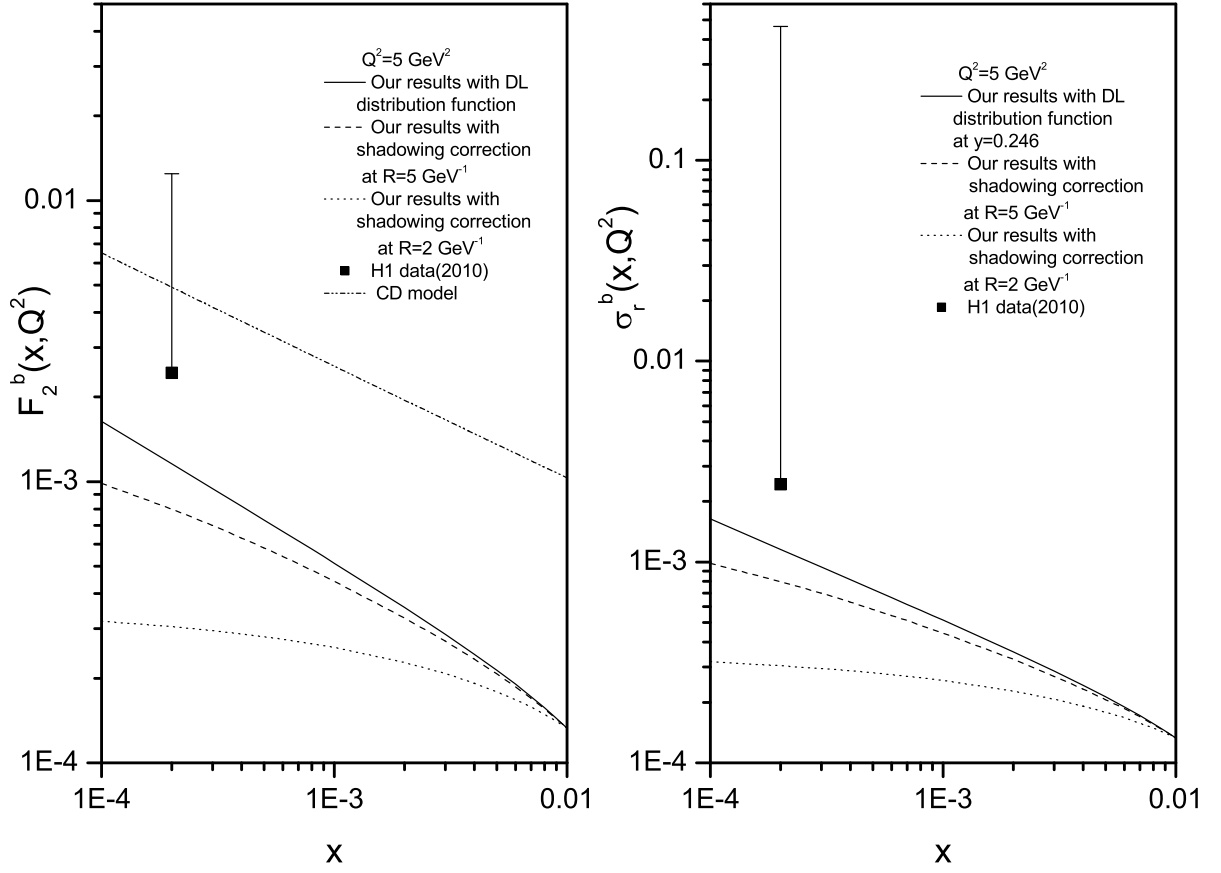}
\caption{The shadowing corrections to the $F_{2}^{b}(x,Q^{2})$
 and $\sigma_{r}^{b}(x,Q^{2})$ at $Q^{2}_{0}=5~GeV^{2}$ from the solution of the initial condition shadowing
 effects with the boundary condition at $x_{0}=0.01$. The solid curve is the
 our results with DL model for unshadowed corrections and dash (dot) curves are the shadowing correction included with
 $R=5~GeV^{-1}(=2~GeV^{-1})$, respectively,
 that compared with CD model (dash-dot-dot) and H1 data. }\label{Fig1}
\end{figure}
\begin{figure}
\includegraphics[width=1\textwidth]{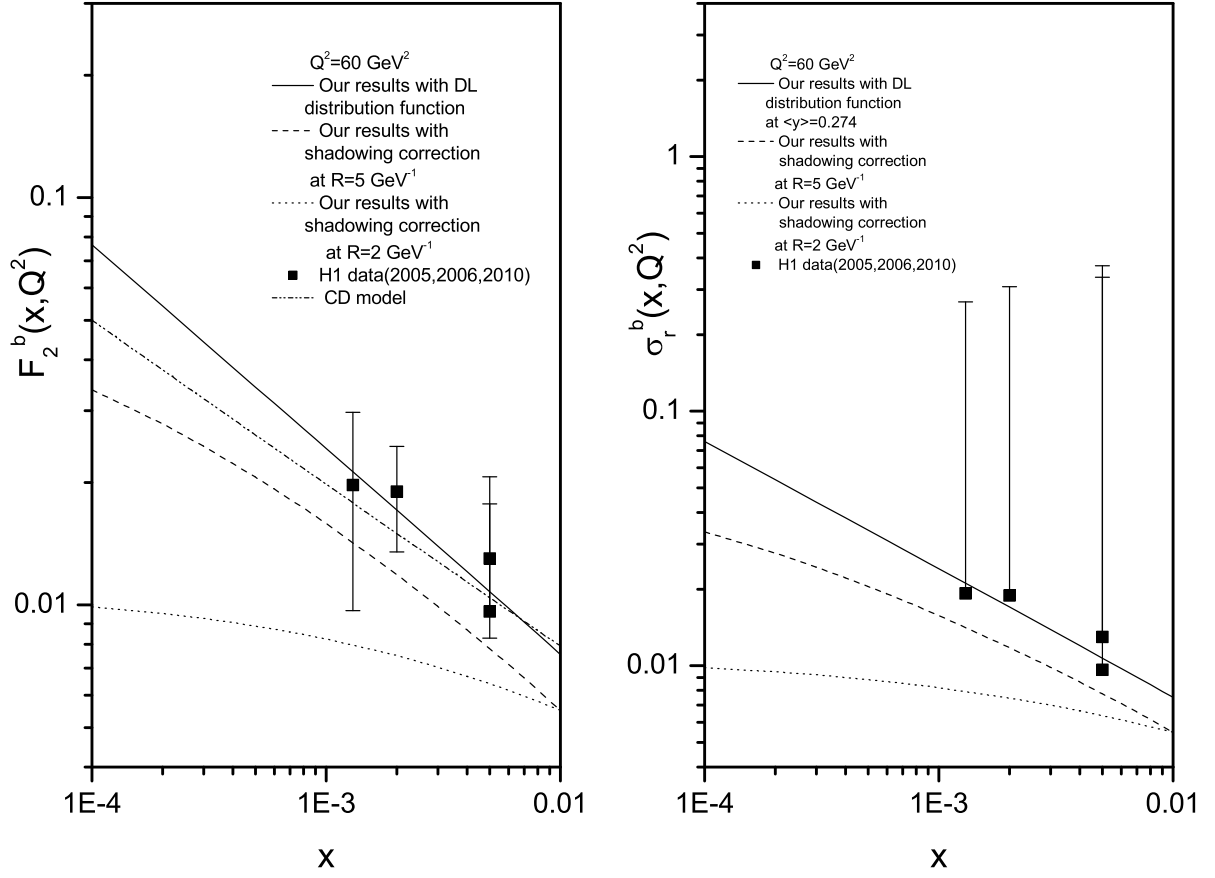}
\caption{The same as Fig.4 at $Q^{2} =60~GeV^{2}$.}\label{Fig1}
\end{figure}
\begin{table}
\centering \caption{Values of $F_{2}^{b}(x,Q^{2})$,
$F_{L}^{b}(x,Q^{2})$ and $\sigma_{r}^{b}(x,Q^{2})$ extracted from
the DL model at low and high $Q^{2}$ (in $GeV^{2}$) values at the
renormalization scales $\mu^{2}=Q^{2}+4m_{b}^{2}$ and
$\mu^{2}=4m_{b}^{2}$, respectively. Also, for comparison, the
results determined by H1 Collaboration (2010)[4] are quoted.
}\label{table:table1}
\begin{minipage}{\linewidth}
\renewcommand{\thefootnote}{\thempfootnote}
\centering
\begin{tabular}{|l|c|c|c|c||c|c|c||c|c|c|} \hline\noalign{\smallskip} $Q^{2}(GeV^{2})$ & $ x$ &
$ y$ & $ {\sigma_{r}}^{b}$ (H1 data )${\pm}  \Delta$ &
$F_{2}^{b}$(H1 data ) ${\pm} \Delta$ & $F_{2}^{b}$ & $F_{L}^{b}$ & ${\sigma_{r}}^{b} $ & $F_{2}^{b}$ & $F_{L}^{b}$ & ${\sigma_{r}}^{b} $ \\
\hline\noalign{\smallskip}
5& 0.00020 & 0.246 & 0.00244${\pm}$0.461 & 0.00244${\pm}$0.01 & 0.00116 & 0.186E-4& 0.00115 & 0.00116 & 0.187E-4 & 0.00116    \\
12& 0.00032 & 0.369 & 0.00487${\pm}$0.318 & 0.00490${\pm}$0.011 & 0.00412 & 0.00015& 0.00411 & 0.00419 & 0.00015 & 0.00418   \\
12& 0.00080 & 0.148 & 0.00247${\pm}$0.435 & 0.00248${\pm}$0.011 & 0.00260 & 0.926E-4& 0.00260 & 0.00264 & 0.94E-4 & 0.00264   \\
25& 0.00050 & 0.492 & 0.01189${\pm}$0.251 & 0.01206${\pm}$0.013 & 0.01040 & 0.000674& 0.01029  & 0.01075 & 0.00070 & 0.01062   \\
25& 0.00130 & 0.189 & 0.00586${\pm}$0.341 & 0.00587${\pm}$0.010 & 0.00645 & 0.000418& 0.00644  & 0.00666 & 0.00043 & 0.00665  \\
60& 0.00130 & 0.454 & 0.01928${\pm}$0.250 & 0.01969${\pm}$0.010 & 0.02124 & 0.002454& 0.02085   & 0.02271 & 0.00263 & 0.02229 \\
60& 0.00500 & 0.118 & 0.00964${\pm}$0.326 & 0.00965${\pm}$0.011 & 0.01078 & 0.001250& 0.01077   & 0.01152 & 0.00134 & 0.01151 \\
200& 0.00500 & 0.394 & 0.02365${\pm}$0.232 & 0.02422${\pm}$0.029 & 0.02488 & 0.00430& 0.02439   &  0.02891 & 0.00506 & 0.02834 \\
200& 0.01300 & 0.151 & 0.01139${\pm}$0.344 & 0.01142${\pm}$0.027 & 0.01528 & 0.00267& 0.01525    & 0.01776 & 0.00314 & 0.01772\\
650& 0.01300 & 0.492 & 0.01331${\pm}$0.347 & 0.01394${\pm}$0.033 & 0.02480 & 0.00430& 0.02397   & 0.03220 & 0.00580 & 0.03109 \\
650& 0.03200 & 0.200 & 0.01018${\pm}$0.301 & 0.01024${\pm}$0.034 & 0.01536 & 0.00274& 0.01530    & 0.01996 & 0.00368 & 0.01987\\
2000& 0.05000 & 0.394 & 0.00499${\pm}$0.611 & 0.00511${\pm}$0.043 & 0.01740 & 0.00288& 0.01710    & 0.02530  & 0.00447 & 0.02480\\

\hline\noalign{\smallskip}
\end{tabular}
\end{minipage}
\end{table}

\end{document}